# Visualizing Digital Collections

Laura Deal[1]

Woodrow Wilson International Center for Scholars

Abstract: Data visualizations can greatly enhance search in digital collections by providing information about the scope and context of a collection and allowing users to more easily browse and explore the contents. This article discusses the benefits of incorporating visualizations into digital collections based on the experiences of the Cold War International History Project (CWIHP) in developing a user-friendly tool for searching and visualizing the project's complex set of historical documents. The paper concludes with a tutorial on using the free Library of Congress tool Viewshare to create visualizations based on real data from the CWIHP Digital Archive.

[1] Laura Deal is a Catalog Specialist at the Woodrow Wilson International Center for Scholars where she leads development of the Cold War International History Project Digital Archive. In 2013 the Digital Archive was awarded the Roy Rosenzweig Prize for Innovation in Digital History by the American Historical Association.

**Visualizing Digital Collections**

Digital collections pose challenges for cultural heritage institutions by upending traditional ways of organizing and accessing their holdings. Many of these challenges overlap, while others are particular to specific kinds of institutions. For archives, it might mean switching to item-by-item descriptions instead of collection-level finding aids. For libraries, it might mean shifting to licensing digital content instead of owning physical objects. For museums, it might mean displaying artifacts as separate digital items without the detailed interpretative framework of an exhibit. For all institutions, the digital shift involves changes to access controls as materials become available beyond the physical confines of brick-and-mortar walls. With this change comes a concurrent loss of control over the usage of digital holdings.

This challenge of organizing and managing digital collections is also a great opportunity to experiment with new ways of displaying, exploring, and understanding cultural heritage materials. For years, institutions have focused on the digitization process itself and the development of the underlying infrastructure necessary to make items accessible online. Now that many institutions have developed the technology and staff expertise necessary to make basic access possible, the next phase will involve leveraging this vast array of digital content in new experimental and potentially revolutionary ways. Increasingly, libraries, archives, and museums have a vast online user-base, patrons whose only interaction with the institution is via a website. Users often come into contact with a digital object first through a search engine, arriving without any contextual knowledge of the collection in which it is contained or the repository that holds it. Providing these users with context and friendly user interfaces is a challenge that cultural heritage institutions can either choose to avoid or approach as an opportunity to create new ways of interacting with and using their digital collections.

Current digital collection interfaces tend to be heavily text-based. Browsing features, if they are offered, are often limited to a list of metadata vocabulary terms or a search that displays the entire contents of a collection, one page of ten to twenty items at a time. The only mechanism for browsing is to flip through thousands of pages of search results and read each individual item. Furthermore, often little to no information is provided about the scope and contents of the digital collection, such as the subject focus, the types of objects it contains, or even the provenance of the items. This is the kind of valuable information that a knowledgeable subject expert can provide through an in-person reference interview. Institutions can try to plug these gaps by providing email or chat-based reference, but it is even more useful to provide basic contextual information about a collection through the website interface itself. One potential solution to this problem of search and providing context in digital collections is the development of data visualizations. Visualizations can be used to show basic features of a collection through easy-to-read graphs and charts, allowing users to quickly grasp the nature of a collection and explore its contents.

## Literature Review

Although computer scientists have been experimenting with data visualization through graphs, charts, and maps for decades, cultural heritage institutions have begun to experiment with visualizations much more recently. As a result, the literature on using data visualization with digital collections is somewhat thin. Most of the studies that do exist are based around a specific use-case, usually a prototype visualization system, and they only occasionally include user-assessment data. Often this prototype is not available for installation and use by other institutions, making the application of new ideas and concepts difficult. Yet, even if the specific tools described by these studies are not readily available, these papers can be helpful for developing new ideas and learning from others' experiences, as well as finding practical advice on designing new data visualizations. To an extent, this article also falls into this case-study genre, although I also provide a tutorial on using one of the simplest currently available free tools—the Library of Congress's Viewshare—to supplement my literature review and discussion of my own project's custom-designed visualizations.

One of the earliest examples of a case study on data visualization is Ahlberg and Shneiderman's classic 1993 description of visualizing data via a “starfield display.” Ahlberg and Shneiderman first explain the then relatively new concept of “visual information seeking” applications and describe foundational concepts for designing systems to facilitate this visual method of search. These concepts include the already well-established “principles of direct manipulation,” with features such as:

- “visual representations of the world of action,” like knobs the users can turn and buttons they can push;
- “rapid, incremental, and reversible actions,” so users can quickly change and adjust their search query;
- “selection by pointing (not typing),” and
- “immediate and continuous display of results” (Ahlberg & Shneiderman, 1993, p. 1).

To this list, Ahlberg and Shneiderman add a number of new principles for supporting visual search:

- “dynamic query filters,” such as “sliders, buttons, etc.” that can be adjusted rapidly by the user;

- use of a "starfield display," which allows all results to be continuously viewed on a single screen, thereby supporting "the viewing of hundreds or thousands of items" at once, and
- "tight coupling" so search refinement options are flexible and interrelated, e.g. "outputs of queries can be easily used as input to produce other queries" (Ahlberg & Shneiderman, 1993).

The titular "starfield display" Ahlberg and Shneiderman (1993) describe is an abstract form of mapping, in which a scatterplot is created based on arbitrary x and y-axes. Using a starfield display, it is possible to represent all available data on a single screen. Each data point on the scatterplot represents an object in the digital collection, such a document, a photograph, or a film (Ahlberg & Shneiderman, 1993). Although their concept of the starfield is not the norm for digital collections today, many of the principles Ahlberg and Shneiderman describe are still a solid foundation for creating useful, user-friendly visualizations. For instance, they recommend that "in a well-designed facility, users should be able to see the impact of each selection while forming a query," so users can quickly make changes and adjust their search parameters based on the automatically updating visual results (p. 3). "Tight coupling" is also a design feature of all modern computing systems, in which, for example, buttons are greyed out when the actions they represent cannot be performed, such as saving when no changes have been made to a document (Ahlberg & Shneiderman, 1993, p. 3). As an example of their design principles, Ahlberg and Shneiderman describe an experimental film database called the "FilmFinder," which displays films on a starfield where "the x-axis represents time [the year of film production] and the y-axis represents a measure of popularity" (Ahlberg & Shneiderman, 1993, p. 4). A user can quickly see the total number of films in the FilmFinder's database and begin limiting results based on filters such as film genre, length, rating, and even specific actors and directors. The resulting system sounds like it might be a much more useful tool for searching and discovering movies than, say, Netflix's atomized lists of film genres.

Much more recently, a team of students at the University of Maryland designed "ArchivesZ," a tool for visualizing archival collections based on Encoded Archival Description (EAD), the XML format which has become the standard for encoding online finding aids (Kramer-Smyth, Nishigaki, & Anglade, 2007). The ArchivesZ prototype interface allows archives users to search for content by year and subject in a tightly coupled dual histogram interface, i.e. two bar graphs in which "as one dimension [date] is manipulated, the other dimension [subject] is updated based on a refinement of collections returned" (Kramer-Smyth et al., 2007, p. 3). Perhaps the most innovative aspect of ArchivesZ is that it uses total linear feet as a unit of measurement rather than the number of separate collections (Kramer-Smyth et al., 2007). This ingenious decision gives users a much better visual representation of the total amount of content available at an archive on a given topic. For instance, a hypothetical archive may own 30 different collections on women's suffrage. On its face, this seems like a very large number until one sees that these 30 collections consist of only 10 linear feet. In contrast, the hypothetical archives owns only two collections related to gender studies, but those two collections may contain more than 100 total linear feet of material. ArchivesZ generates its subject list by breaking down Library of Congress Subject Headings into component tags, so the long heading "Tobacco—Maryland—History" becomes the list of separate tags "tobacco," "Maryland," and "history." (Kramer-Smyth et al., 2007, p. 7). In this way, all collections about tobacco, Maryland, or history can be grouped together, rather than displaying the single collection tagged with "Tobacco—Maryland—History" by itself. The code for the ArchivesZ prototype is available online along with instructions for installing it—which requires a MySQL database and an installation of Ruby on Rails (Anglade, 2007). Additionally, one of the original team members, Jeanne Kramer-Smyth, has a blog with many examples of experimenting with ArchivesZ using data from real EAD finding aids donated by a variety of different archives (Kramer-Smyth, n.d.).

A number of other data visualization prototypes have been designed in recent years. These include a visual interface for the Internet Archives' “Archive-It” tool developed by Kalpesh Padia for his thesis at Old Dominion University (2012). As Padia explains, “Archive-It is a web archiving service that allows individuals and organizations…to create and archive collections of web pages” (2012, p. 1). Padia developed a variety of visual interfaces for exploring Archive-It collections, including a treemap (a hierarchical chart which shows proportional information using nested rectangles), a “time cloud” (a word cloud that incorporates temporal information), a bubble chart, an image plot, a timeline, and a Wordle tag cloud (2012). Another team from the Netherlands, Scharnhorst, ten Bosch, and Doorn (2012), designed a visual interface for the self-archiving system “EASY” at the Data Archiving and Networked Services (DANS). DANS “is the largest national research data archive in the Netherlands in the social sciences and humanities” and EASY is the institution's digital archive for research data sets (Scharnhorst, ten Bosch, & Doorn, 2012, p. 3). The team created a set of interactive web-based visual interfaces for use in navigating EASY's collections using an interactive treemap and a hierarchical category tree. In another study, a US-Canadian team, Ruecker, Shiri and Fiorentino (2012), developed two different thesauri interfaces with visual elements, “Searchling” and “T-Saurus.” Searchling was designed to make complex thesaurus hierarchies and bilingual dictionaries easy to browse, while T-Saurus represents the number of terms found by a query through a visual display with different sets of "buckets." In T-Saurus, "the number of buckets represent the number of terms found by the query," and "the size of the buckets represents the number of matches for that particular term, while proximity and opacity represent scope and accuracy of the term in relation to pre-established hierarchies" (Ruecker et al., 2012, p. 3). These prototype interfaces may be useful as examples for anyone developing a hierarchical browser. Furthermore, a slightly different but related team, Shiri, Ruecker, and Murphy (2012), compared the two different thesauri interfaces in user evaluations in order to draw conclusions about the search preferences of "linear" thinkers vs. "visual" thinkers. Lastly, Mark Hall and Paul Clough of University of Sheffield (2013) describe an original map-based visualization for exploring large document collections. In their highly technical paper, Hall and Clough explain how they created a Hierarchical Spatialization Algorithm, which can be used to create "a hierarchical, semantic map" representing the extent of different topics contained in a single document set (2013, p. 3).

Aside from these more focused case studies, a number of authors have also begun writing more abstractly on the philosophy and reasoning behind incorporating data visualizations into digital collections. Mitchell Whitelaw (2012b), for example, has written extensively on the theory and practice of visualizing digital collections as well as the design of user-friendly "generous interfaces," which incorporate visual elements. Whitelaw also notes that the time is ripe to develop new interfaces based on the assumption that many users' "only experience of a collection will be digital" and thus the website interface becomes "the manifestation of that collection for the users" (2012b, p. 1). In his introductory article on generous interfaces, Whitelaw (2012b) begins by describing current interfaces for digital collections, which almost universally place primacy on text search (p. 1). New users are regularly faced with a blank set of search boxes and little to no information about the contents of the collection they are searching. Search becomes a series of trial-and-error tests that too often result in a disappointing black page with “no results found.” Text search assumes that users know what they are looking for and that they do not need assistance exploring the collection, yet, as any reference professional, knows this is rarely the case. In contrast, Whitelaw's “generous interfaces” do not hide their inner treasures; they show them to the user from the very beginning, giving users an overview of the content before they search. Whitelaw has created several prototype projects based on these principles for Australian institutions, including Manly Images, “an experimental web interface to the Manly Local Studies Image collection,” (Whitelaw, 2012a) and the commonsExplorer, “an experimental interactive browser for Flickr Commons” (Whitelaw & Hinton, 2010).

In their article on visual content exploration in large document collections, Drahomira Herrmannova and Petr Knoth (2012) speak on similar themes when they explain that “while exploratory searches constitute a significant proportion of all searches, current search interfaces do not sufficiently support them” (Introduction, para. 3). Herrmannova and Knoth then go on to describe the variety of new approaches that have been developed for visualizing digital collections, outlining basic design principles such as “added value” (visualizations should “reduce the mental workload of the user,” not simply look cool); a focus on simplicity in design and high “visual legibility” to reduce new users' learning curve, as well as the thoughtful use of colors, dimension, and fixed spatial location in visualizations (2012, Section 3, para. 3-8). Herrmannova and Knoth then propose a hypothetical visual interface for better supporting exploratory search which consists of a screen divided into three columns, including a central “visualization area” and “a left and right sidebar” (2012, Section 4.3, para. 1). Users may perform a text search on the left sidebar and see a list of results there, while details of a selected item are displayed in the right sidebar. Users can then drag a specific item into the central visualization area, which shows connections between documents along pre-determined dimensions—such as links formed between documents by the same author or documents about the same topic (Herrmannova & Knoth, 2012, Section 4.3.1, para. 1). In this way “stacks” of related documents can be formed and even more connections explored between related documents.

Wilko van Hoek and Philipp Mayr (2013) also provide a broad overview of visualization techniques for use in digital libraries. Van Hoek and Mayr developed their own useful definition of search itself, dividing the search process into three distinct parts: forming a query, browsing results, and close examination of a specific item (2013). They then provide advice and examples for supporting every stage of the search process through helpful visualizations. This breakdown of the search process will be useful for institutions that are developing new visual interfaces as van Hoek and Mayr help clarify how each different type of visualization is able to support the complex three-step search process.

Another article worth mentioning is Alex Byrne's abstract exploration of the challenges of representing digital and physical collections to enable discovery and use (2012). Writing from the perspective of a director of a large state library, Byrne discusses broadly the visual nature of collections, whether digital or analog, and questions whether it is even possible to visualize them as one complete unit:

> How can we mentally "see" collections spread between open access shelving, traditional stacks, dense storage, and automated systems? And, how can we meaningfully comprehend their cultural and informational content across the myriad formats and topics represented in extensive collections? (p. 16)

Byrne then discusses the specific limitations of library stack organization and bibliographical catalogs, including new social catalogs like Library Thing and Goodreads (2012).

Lastly, another source on the theory of visualization is “Data Stories,” a bi-weekly podcast on data visualization hosted by Enrico Bertini and Moritz Stefaner (2012). Many of the same concepts and people cited above have been discussed or have even appeared as guests on Data Stories. Although using an ironically un-visual medium, this podcast is an excellent source of information on new concepts and projects experimenting with data visualization. Each episode is also nicely indexed into useful chunks on the blog.

## The Future of Visualizations

Cultural heritage institutions have experimented with visualizations for digital collections since at least the 1990s, yet they have not become a common feature of most search interfaces. Why have they so far failed to become widespread? In their assessment of visualization techniques for search in digital libraries, Wilko van Hoek and Philipp Mayr (2013) theorize that it is because users are still largely unfamiliar with visual interfaces and therefore perform poorly when first learning to use them during user evaluations. Users perform better on text-based search because they have more experience with text-based interfaces. In contrast, users are not given sufficient time to learn new visual-based interfaces in the short time measured by most assessment studies. Furthermore, van Hoek and Mayr argue that the available studies are too varied to draw definitive conclusions about the usefulness of visualization techniques, but note the high level of user satisfaction with and interest in visual-based search interfaces (2013). Given the limited number of studies that even attempt to assess visual interfaces, another, more likely explanation for the slow adoption of data visualizations for digital collections is the large initial cost required to develop a unique new interface. Building prototypes is an expensive, time-consuming process that only a few institutions have the resources to undertake. The prototypes that have been developed often do not become incorporated into finished digital libraries and remain only intriguing curiosities or proofs of concept. Despite this slow start, the time seems ripe now for creating visualizations of digital collections.

Scharnhorst, ten Bosch, & Doorn (2012) note that “the last decade has also seen a movement to popularizing and democratizing visualization methods” (p. 2). Infographics have become incredibly popular and widespread online thanks to the format’s ability to convey a clear message and explain complex concepts with simple graphs and charts. The rise of infographics stems in part from the increasing availability of large research data sets and useful tools for manipulating and displaying data like OpenRefine, Google Charts, IBM Many Eyes, and Freebase. Similarly, cultural heritage institutions have laid the foundation for new innovative uses of data through the mass digitization of large collections of material. The new Digital Public Library of America (DPLA), for example, is building on this huge amount of digitized content by aggregating material from partners with large collections of more than 250,000 unique digital items. Already the DPLA has 5 million objects and counting, having collected only a very small subset of the total digitized content available online from American institutions alone (Digital Public Library of America, 2013).

The DPLA has also innovated by opening up its data through an Application Programming Interface, or API. Through APIs institutions can expose their data and allow others to build new tools based on the huge amount of useful cataloging and descriptive metadata that has already been created. As Daniel J. Cohen (2006) explains,

> APIs often include complex methods drawn from programming languages—precise ways of choosing materials to extract, methods to generate statistics, ways of searching,

> culling, and pulling together disparate data—that enable outside users to develop their own tools or information resources based on the work of others (Lessons section, para. 2).

Users may be familiar with API-based tools such as the proliferation of apps based on the Twitter API. These apps are able to extract tweets, hashtags, and data directly from Twitter. This proliferation of creative third-party tools is made possible by APIs, which, in Cohen's words, "provide fertile ground for thousands of developers to experiment with the tremendous indices, services, and document caches maintained by" for-profit companies (2006, Lessons section, para. 4). Cohen goes on to explain that "owners of digital collections can create a rudimentary API by repackaging their collection's existing search tool using simple web services protocols" (2006, Lessons section, para. 5). The CWIHP Digital Archive has a simple API that allows users to extract data in a variety of formats, including XHTML, JSON, YAML, XML and CSV.

While many cultural institutions may not have the money, expertise, or time to create experimental visualization tools, there is a growing collection of communities that do, including researchers in the digital humanities, citizen scientists, and crowdsourcing volunteers. The Digital Public Library of America's API has already resulted in a host of new apps based on the aggregated content of its dozens of cultural institutions. APIs and outsider apps take the burden of innovation off the shoulders of individual institutions and open up cultural heritage collections to new creative experiments that might never have attempted otherwise. Of course, APIs also require letting others play in your sandbox and losing complete control over your own content—something libraries, archives, and museums sometimes struggle to accept. Many institutions purposefully keep data locked down and perform all web development internally to avoid this inherent loss of control. Both strategies, either opening up data to allow others to experiment or creating everything in-house, have risks. In-house development is expensive and resource-heavy, and may lead to staff overload and project stagnation, while opening up data could result in outside users making use of collections in ways that institutions do not like. Cultural institutions must each decide individually if the potential benefits are worth the hazards.

Open data also gives institutions the possibility of using open-source plugins based on common content management platforms and metadata standards. For example, the Library of Congress's Viewshare tool can import data through a variety of means, including OAI-PMH and some versions of ContentDM. Viewshare can be used as a stand-alone platform for displaying a digital collection, or as a way of adding value to a pre-existing collection website. Users can create new "views" or visualizations for collections, thereby providing new ways of exploring and understanding the underlying content. Viewshare is built on Exhibit, a framework for creating interactive visualization from MIT. Exhibit is a flexible set of tools that can also be used to make stand-alone web pages, but it is a bit intimidating to set up and requires some basic coding knowledge (SIMILE Widgets, n.d.). Other examples of free visualization tools include Vidi, a set of data-visualization modules developed by the Jefferson Institute for Drupal (Jefferson Institute, 2010); Neatline, a suite of map and timeline tools developed for the open-source Omeka content management system (Scholars' Lab, n.d.); and Elastic Lists, an open-source facet browser developed by Moritz Stefaner which allows users to visually filter results based on metadata categories (Stefaner, n.d.). New tools proliferate every day, some of which are quickly adopted, while others fall into obsolescence. Open-source tools have the benefit of being free, but they may also require a high degree of technical knowledge to install and use successfully. Below, I have chosen to focus on Viewshare for my tutorial because it is a very easy-to-use tool that was specifically designed to quickly and easily import data from existing digital collections.

## Types of Visualizations

In an effort to condense the many different types of visualizations, I have categorized them based upon the three different overlapping functions they serve for search and discovery:

### Descriptive

Descriptive visualizations give users contextual information about a collection before they begin searching. This upends the more common interface design that presents a user with an empty search box and little to no information about the collection within. As librarians and archivists are well aware, the best searchers are people who already have good knowledge of the collections they are searching. As Hinton and Whitelaw (2010) note, “search assumes that a user is able to provide a query; but a user who is unfamiliar with the collection’s scope, contents, or structure may not be in a position to query it effective” (p. 52). When faced with a contextless search box, search becomes a process of trial-and-error, often with many dead-end queries and a frustrating lack of results. This can be avoided by providing a graphical representation of the collection that helps the user immediately see its scope and contents. Drawn from metadata, visualizations can reveal a collection’s composition such as date range, languages, types of material, common subjects, and creators. They might include a pie chart, bar graph, map, or other useful charts or graphs depending upon the nature of the underlying collection and the available metadata.

### Browsing

Purely descriptive visualizations provide users with context to assist them with a text-based search. In contrast, browsable visualizations can be manipulated by the users to perform a search based on the visual information with which they are presented. This interactivity introduces an element of play, allowing users to experiment and explore the collections, in the process revealing new aspects of the content and potentially unexpected features that are not evident in traditional forms of description. It also makes search more fruitful, as Ahlberg and Shneiderman (1993) explain, "in a well-designed facility, users should be able to see the impact of each selection while forming a query…. The idea is to prevent users from specifying null sets" and hitting the brick wall of 0 results found (p. 246).

### Visual search options

Separate from data visualization per se, but significant for ease of user search is the creation of visual search mechanisms in the interface itself. These are "visual representations of the world of action" also described by Ahlberg and Shneiderman (1993), that is, any visual means of simplifying the selection of search options, such as a calendar interface for selecting dates or a map for selecting geographic locations (p. 1). Visual search mechanisms first show the user the underlying structure of the data and then have them select an option, rather than requiring the user to use trial-and-error searches to understand the structure and limitations of the system.

## Digital Collections and Iterative Development

Digitizing and cataloging a collection is by no means a small feat, and I do not wish to disparage current limited digital interfaces or the time and funding cultural institutions have invested in digitizing their collections. Simply making materials available in a searchable digital format is a huge first step, and one that deserves congratulations. Yet it is best thought of as only a first step. In order to stay current and useful, digital collections should continue growing and changing as new technology develops and institutions' abilities and priorities shift. Each new step and new upgrade builds a solid foundation for the next step and even greater improvements in the future.

My own digital project is a good example of this long-term iterative process of development. The Cold War International History Project (CWIHP) was founded in 1991 in response to the fall of the Soviet Union and the opening of former-Soviet archives in Eastern Europe. The project began collecting copies of Cold War-era historical documents from these newly opened archives and encouraged Western historians to use Eastern sources. Since then, the project has served as a document clearinghouse, accepting copies of archival collections from researchers as well as gathering copies of documents directly from the archives themselves. The CWIHP archive has grown over the past 20 years through both formal and informal means, becoming in some cases the only place researchers can access documents that have been re-classified in their country of origin.

In 1995, the project built its first website and began posting material online (Hershberg, 1995, p. 125). This early website had a “Cold War Virtual Library,” which included English translations of historical documents made available with rudimentary metadata and basic HTML formatting. The Virtual Archive went through several iterations as the Cold War project’s website moved, first hosted by George Washington University, then the Smithsonian Institute, and finally appearing on the Wilson Center’s own dedicated website. In 2003, the Virtual Archive was re-launched as part of the Wilson Center’s Cold Fusion-based website, and the metadata scheme was revised to match the Dublin Core standard for digital assets. A Google search plugin on the website offered basic full-text search.

In 2011, thanks to a series of digitization grants and internal support from the Wilson Center itself, the project was able to begin planning for the complete redesign and re-launch of the CWIHP website, including a renamed Digital Archive. After a competitive bidding process, Second Story Interactive Studios was selected as the project’s web developer due to its experience working on museum exhibits and digital collections as well as the elegance and interactivity of their designs. One of the project's key goals was the development of user-friendly browse tools that would make it easy and fun for new users to explore the diverse collections of the Digital Archive. The unique nature of our collections— drawn from more than 100 different archives in dozens of different countries around the globe—make them particularly challenging for new users to explore.

## The New Cold War Digital Archive

The Cold War project’s forays into data visualization began with the third function of visualizations, visualizing search options. The Digital Archive’s previous interface allowed only browsing across artificially created thematic “collections,” or an alphabetical list of geographic locations. These geographic terms refer to the subject content of each individual document, so selecting “Czechoslovakia” leads the user to a list of historical documents about the former country, as opposed to documents from Czechoslovakia. This is in part because the originating location of a document often says very little about its subject content and in fact can be quite misleading since much of the Digital Archive focuses on diplomatic material in which representatives from one country describe and analyze events in another. For instance, while the Digital Archive contains no documents from North Korea (aside from a handful of captured records from the Korean War), it has nearly 1,500 historical documents about the country drawn from the archives of North Korea’s former communist allies.

Geography is one of the most significant ways of accessing the Digital Archive since Cold War scholars often focus on a specific region or country. Similarly, geographic browsing is an easy entry point for new users who are unsure what is contained in the collections and simply want to explore. One of the goals of the new website's development was to design a map interface that would simplify the selection of geographic regions and countries for search (Digital Archive, n.d.-a) (see Figure 1). The new interface was designed to support a hierarchy, allowing countries to be associated with a larger region, and cities with a larger country. On the map itself, the geographic regions would be displayed as simple outlines without internally defined political boundaries due to the complications of displaying the shifting borders before and after the Cold War.

**Figure 1**: Digital Archive "browse" page, with East Asia selected.

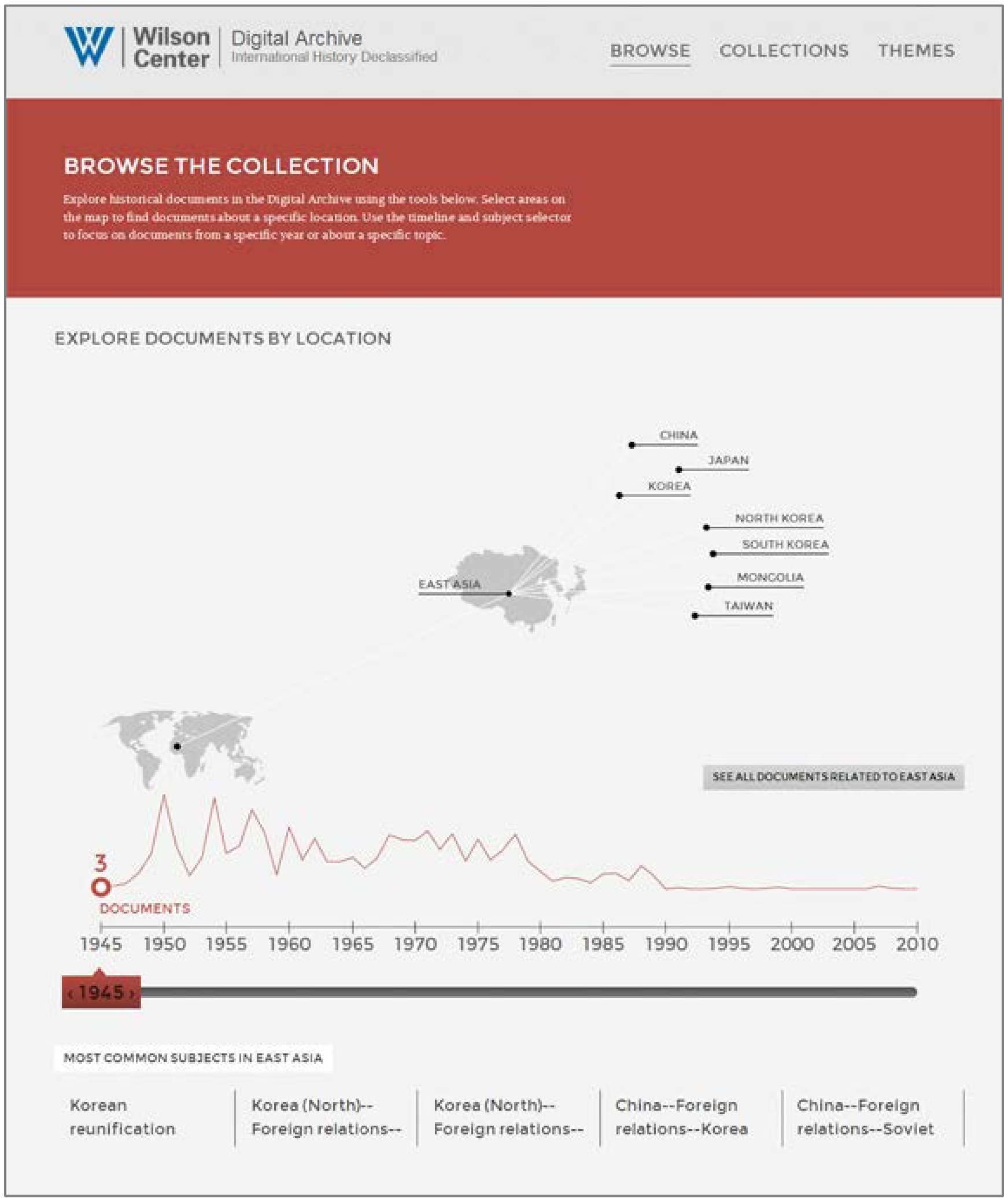

To expand the map beyond a simple geographical search interface, the project also incorporated descriptive information about the collections and interactive browsing features. Below the map, a timeline displaying the number of documents by year gives the user an immediate overview of the date range of documents in the Digital Archives and basic features of the total collection (Digital Archive, n.d.-a). For instance, the user can immediately see that 1962 contains the most documents by far, representing a huge set of documents released by the project in October 2012 for the 50th anniversary of the Cuban Missile Crisis (see Figure 1). Below the timeline, the five most common subject headings are displayed, giving the user basic subject access to the collections. The map is also browsable and interactive. When a user selects a specific region, country, or city, the timeline and list of subjects below is automatically updated, quickly showing the strengths and weaknesses of document coverage in each geographic area. Clicking on “East Asia,” for instance, shows that most of the Digital Archive’s relevant documents are concentrated between 1950 and 1975, and the most common subjects all focus on North Korea and China’s foreign relations (see Figure 1).

At this current stage, the browse tools are very simple and represent only a first, experimental step toward visualizing our collections. The timeline and subject list provide only the simplest overview of the Digital Archive’s collections, and currently selecting any of the options (a geographic region, a year, or a subject) sends the user to a list of search results that they must page through one by one to continue browsing. Ideas for future updates include adding visual filtering tools to the search-results page itself, adding visualizations that show the original languages of each collection’s documents, and improving the subject list display.

In the 8 months since the Digital Archive's re-launch, the new browse-page visualizations have shown robust usage. Browse is the third most popular page on the Digital Archive website, following the front page and the list of thematic "collections." The bounce rate (the ratio of users who exit the site immediately upon seeing the browse page) is relatively low, with 48% of users clicking on an item to continue their exploration of the Digital Archive. Users spend an average of 44 seconds on the page, as compared with the Digital Archive's overall time-on-page average of 1:11. Initial feedback has also been good. The site was launched first with a small beta group of 26 users who provided detailed feedback on the new design, including the browse page. The beta testers were drawn from CWIHP's current user base and included people with different levels of technical expertise. They included students (25%), college professors (20.8%), independent scholars (37.5%), librarian/archivists (8.3%), and government officials (8.33%). Of respondents, 15.4% had never used the Digital Archive website before, while 84.6% had used it at least once previously. Several users highlighted the browse page for praise, with comments such as "I like… the fact that you can focus on a particular geographical area" and "a very creative set up... good navigation and user-friendly." Overall comments stressed the ease-of-use and appealing look of the new website, with 69.23% of those who had used the previous Digital Archive rating the new design as "significantly better" than the older, highly text-based interface. In the future, a larger survey is planned in order to get more feedback from users before new features are implemented such as new visualizations and user accounts.

## Tutorial – Experimenting with Viewshare

As an example of the kind of visualizations that can be developed using freely available tools, I created three different "views" on Viewshare using data drawn from the CWIHP Digital Archive. These views and the underlying data can all be accessed via my profile on Viewshare, username lauradeal (Deal, 2014b). For the most part, I found the Viewshare interface very intuitive and easy to use. There is a detailed user guide available to help new users get started and the @ndiipp twitter account run by Viewshare's administrating program, the National Digital Information Infrastructure and Preservation Program (NDIIPP), is generally very responsive to questions (Viewshare, n.d.-b). Accounts must be requested but are "available to individuals associated with cultural heritage organizations" including "libraries, archives, museums, historical societies, colleges and universities" (Viewshare, n.d.-a).

Viewshare has a large number of importation options, including pulling data directly from a JSON list, an OAI-PMH endpoint, XML MODS files, or a ContentDM database. If your data is not exposed through any of those protocols, or you are unfamiliar with using web services, you can also import data from a simple CSV or Excel spreadsheet. Importing data directly through JSON, XML, or OAI has the benefit that the data updates automatically and any "views" created from that dataset will remain up to date. However, your ability to edit and transform such data is limited by Viewshare's interface options. In contrast, importing from a CSV or Excel spreadsheet gives you the ability to easily transform and combine data in Excel prior to importation into Viewshare. But this data must be refreshed manually by reimporting/uploading the spreadsheet file. Importing via web services is most useful for collections that are updated regularly and are constantly changing, while importing via spreadsheet is most useful for closed collections that will not be updated again or are only rarely updated.

For my own example, I imported data both directly from the Digital Archive's JSON web services and also from Excel files (Digital Archive, n.d.-b). I found that using spreadsheets gave me much more flexibility in transforming and editing data but required more preparation time than simply importing directly from the web. For example, metadata about whether or not a translation of a document is available is stored as a checkbox in the Digital Archive content management system, appearing as either "checked" or "unchecked" in the imported metadata. By starting with a simple spreadsheet, I was able to quickly find/replace these terms for the more user-friendly "Translation Available" or "No Translation" (see Figure 2). Spreadsheet data can be easily re-uploaded or "refreshed" after edits have been made to the original file. After some trial and error, I also discovered that Viewshare had an easier time importing Excel files than plain CSV files. The metadata fields I imported for my three data sets included document title, document description, original language, document date, geographical location, translation status, record ID, and URL.

**Figure 2**: Metadata refinement after importing a spreadsheet into Viewshare.

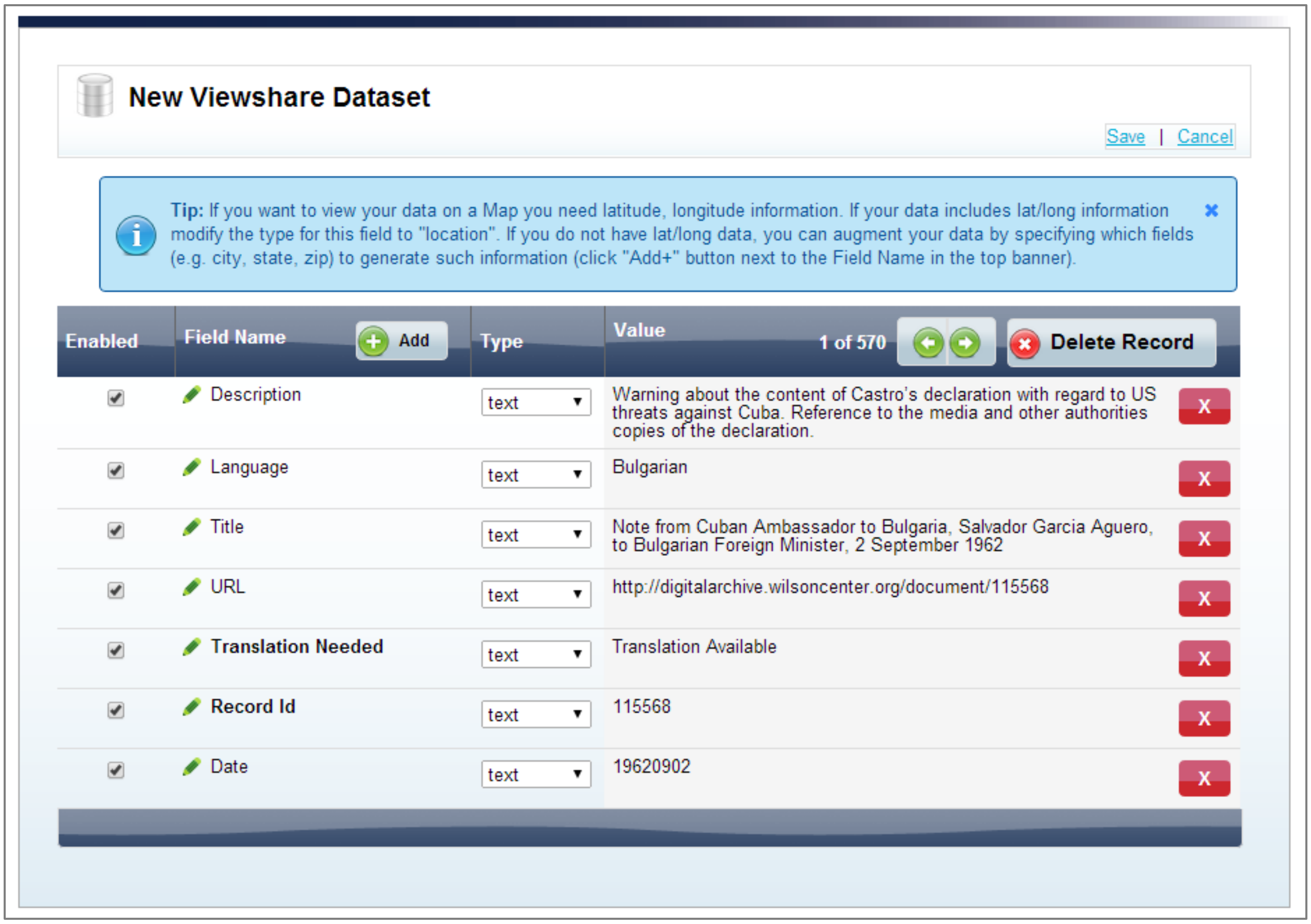

Once the data has been uploaded, Viewshare displays an example record with each of the metadata fields it has detected (see Figure 2). Viewshare's interface allows you to assign data "types" to each field in the metadata scheme, sorting the data into categories such as text, number, date/time, location, and URL. This basic administrative metadata helps Viewshare interpret your data and create specific kinds of graphs and charts later in the visualization process. For example, a text field can become a list, a date field can become a timeline, and a location field can become a map. Data can also be edited and transformed in limited ways via Viewshare's interface. Viewshare has the ability to "augment" certain fields and reformat data to better work with its interface. Currently you can augment only a date, location, or list field. Using these options, Viewshare can transform dates into ISO 8601 date format (YYYY-MM-DDT00:00:00+00:00), transform city/state names and zip codes into exact latitude/longitude (Washington, DC becomes 38.89511,-77.03637), and split text into a list using preset patterns or delimiters ("bread, flour, milk" or "bread; flour; milk" or "bread--flour—milk," etc.). Two or more fields can also be combined into one using augmentation.

Once your data has been edited, you can save the data set and begin building a "view" or data visualization (see Figure 3). Multiple views can be made from the same dataset, so you can create several different ways of visualizing the same collection, such as for different audiences. Note that multiple data streams cannot currently be combined and used for the same view. Only one data set at a time can be used for each view. Views are created by selecting from a variety of options, such as creating a pie chart, a map, a scatter plot, a table, a timeline, or a list. "Widgets" can also be added to the sides or top of the view, including a simple search box, a list for filtering, a number slider, a tag cloud, or your institution's logo. There are a whole host of different possibilities and flexible options for displaying data. It takes some experimentation to determine what all of the different dropdown boxes and checkboxes do, but overall it is very easy and fun to create a new visualization using Viewshare.

**Figure 3**: Creating a "view" using Viewshare.

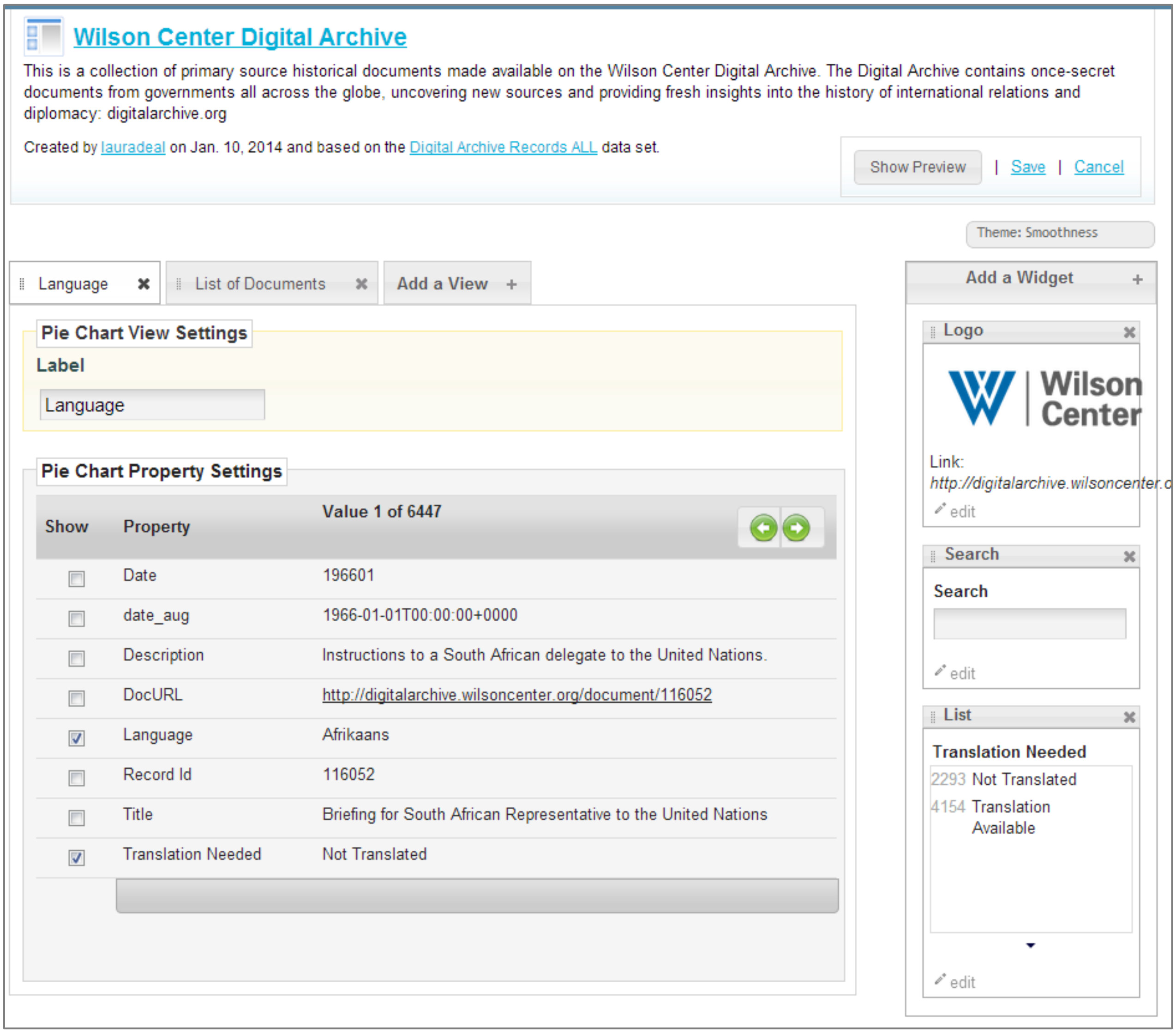

The three views I created myself included a timeline using a sample of the first 500 records imported into Viewshare via JSON (Deal, 2014c), a pie chart showing all of the Digital Archive's documents sorted by original language (Deal, 2014d), and a detailed view with several different options for exploring our Cuban Missile Crisis collection (Deal, 2014a). Overall I was very pleased with my ability to create interesting and informative views. The pie chart displaying documents by original language is a particularly useful visualization I would like to incorporate into the Digital Archive's "browse" page in the future (see Figure 4). Searching in Viewshare is also a pleasure. Search is very quick, and visualizations are automatically updated to show the resulting changes as the displayed data is narrowed based on search terms. For example, on the language view, searching for "Khrushchev" updates the pie chart, revealing that of the available documents that mention Khrushchev, 55.8% are Russian, 8.8% are German, 5.5% are Albanian, 5.5% are Polish, 5% are Romanian, 3.9% are Czech, 2.8% are Chinese, and so on. Using the translation field "widget" I created on the right side of the view, a user can quickly limit documents to only those with English translations available or only those without translations (see Figure 4).

**Figure 4**. [Viewshare pie chart showing the percentage of documents by language.]

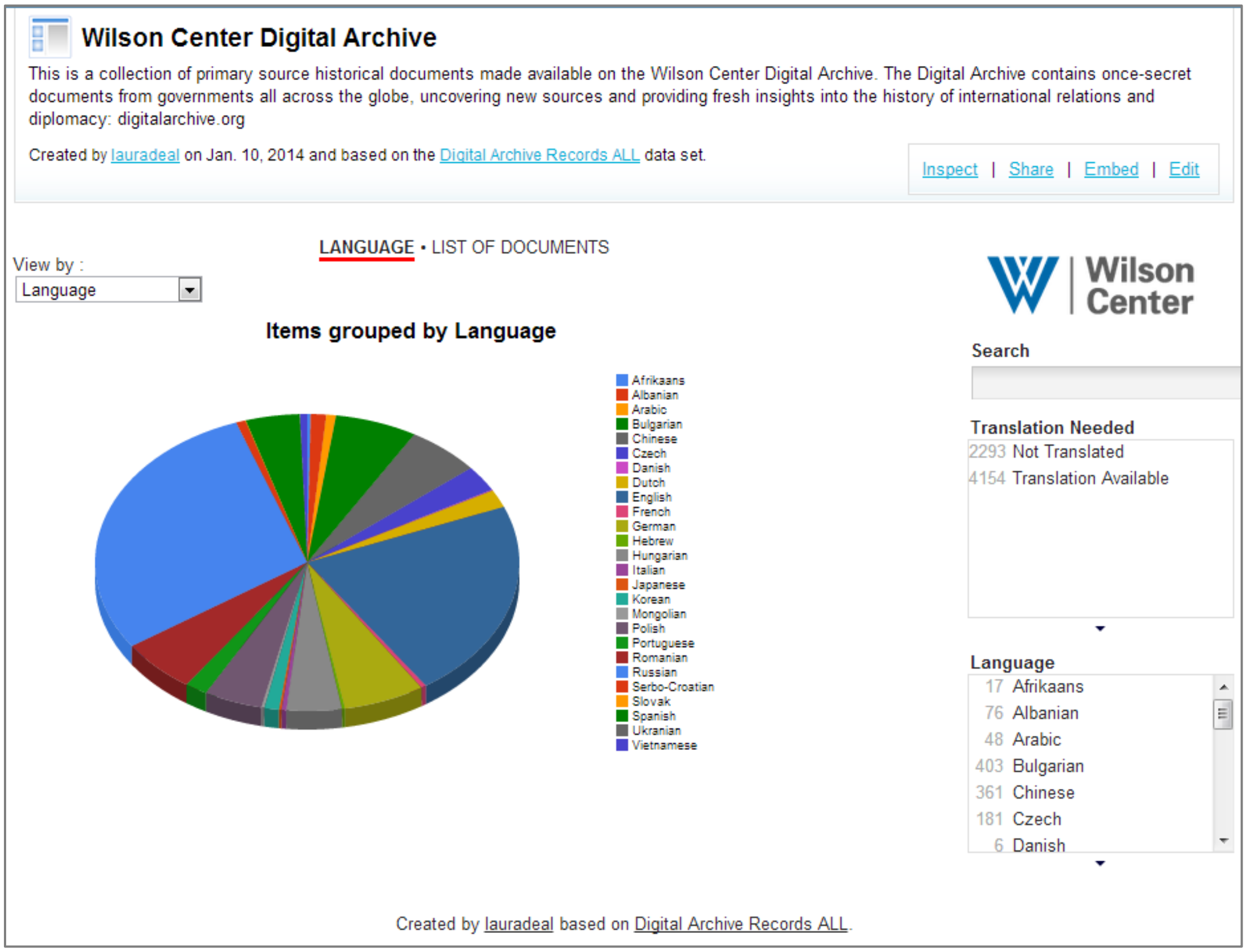

Once created, Viewshare views can be embedded directly into webpages, making it possible for an institution to embed them directly into a digital collection website. Views could also be useful for displaying collections on a blog, a research guide, a finding aid, or an online catalog. The flexibility and easy-to-learn interface of Viewshare makes it an excellent option for experimenting with visualizing digital collections. I strongly recommend that anyone with a sample data set try using it and test out the variety of options. The potential uses and possibilities are nearly limitless.

## Conclusion

The digital landscape is changing rapidly, bringing new opportunities for cultural heritage institutions to update and innovate with their digital collections. Visualizations have the potential to greatly improve search and discovery for online collections, transforming how users interact with digital collections. Furthermore, changing technology is making it easier than ever to incorporate visualizations into search interfaces and websites. The time is ripe for cultural heritage institutions to begin experimenting with data visualization in earnest. The first step is simply testing freely available online tools and exploring the endless possibilities of data visualization for digital collections.